\documentclass[fleqn,10pt]{wlscirep}
\title{Stokes flow around an obstacle in viscous two-dimensional electron liquid}

\author[1,*]{G. M. Gusev}
\author[2]{A. S. Jaroshevich}
\author[1]{A. D. Levin}
\author[2,3]{ Z. D. Kvon}
\author[2]{A. K. Bakarov}
\affil[1]{Instituto de F\'{\i}sica da Universidade de S\~ao
Paulo, 135960-170, S\~ao Paulo, SP, Brazil}
\affil[2]{Institute of Semiconductor Physics, Novosibirsk
630090, Russia}
\affil[3]{Novosibirsk State University, Novosibirsk 630090,
Russia}

\affil[*]{gusev@if.usp.br}

\begin{abstract}

The electronic analog of the Poiseuille flow is the transport in a narrow channel with disordered edges that scatter electrons in a diffuse way.
In the hydrodynamic regime, the resistivity decreases with temperature, referred to as the Gurzhi effect, distinct from conventional Ohmic behaviour.
We studied experimentally an electronic analog of the Stokes flow around a disc immersed in a two-dimensional viscous liquid.
The circle obstacle results in an additive contribution to resistivity. If specular boundary conditions apply, it is no
longer possible to detect Poiseuille type flow and the Gurzhi effect. However, in flow
through a channel with a circular obstacle, the resistivity decreases with temperature.
By tuning the temperature, we observed the transport signatures of the ballistic
 and hydrodynamic regimes on the length scale of disc size. Our experimental results confirm theoretical predictions.

\end{abstract}

\begin{document}

\flushbottom
\maketitle
%
%
\thispagestyle{empty}

\section*{Introduction}

In the absence of disorder, an interacting many-body electron system can be described within the hydrodynamic framework \cite{andreev, principi, lucas}.
Typical three-dimensional metals rarely enter into the hydrodynamic regime because the electron-impurity (phonon) scattering
is stronger than the corresponding electron-electron interactions \cite{lucas1}. However, it is expected that in a clean two-dimensional
(2D) electron system, such as modulation doped GaAs systems and high-quality graphene layers, the requirements for hydrodynamics
can easily be satisfied.

Hydrodynamic characteristics are enhanced in a Poiseuille geometry, where a parabolic flow profile can be realized in a narrow pipe. The fluid in this regime has zero velocity at the boundaries.
The electronic analog of the viscous flow in the pipe is a transport in a narrow channel of width $W$ with diffusive scattering  at the boundary, driven by the electric field.
Viscous electron flows are expected to occur when
the mean free path for electron-electron collision, $l_{ee}$, is much shorter than the sample width,
while the mean free path due to impurity and phonon scattering, $l$, is larger than $W$. It has been predicted that the electrical resistivity of a 2D system is proportional
to electron shear viscosity, $\eta=\frac{1}{4}v_{F}^{2}\tau_{ee}$, where $v_{F}$ is the Fermi velocity and $\tau_{ee}$ is the
electron-electron scattering time $\tau_{ee}=l_{ee}/v_{F}$ \cite{gurzhi, dyakonov, dyakonov1, dyakonov2, govorov}.
 For example, resistance decreases with the square of temperature,
$\rho \sim \eta \sim \tau_{ee} \sim T^{-2}$, referred to as the Gurzhi effect, and with the square of sample width
$\rho \sim W^{-2}$. The boundary conditions can be characterized by a diffusive scattering or by a slip length $l_{s}$ with extreme cases being
no-slip ($l_{s} \rightarrow 0$) and no-stress ($l_{s} \rightarrow \infty$) conditions. It is expected that for $l_{s} \rightarrow \infty$ no
Gurzhi effect should be detected.

Recently interest in electronic hydrodynamics has arisen from measurements of the transport in graphene, where electron-phonon scattering is relatively weak
\cite{levitov, torre, pellegrino, bandurin}. Moreover, a series of updated theoretical approaches has been published \cite{alekseev, scaffidi, luca, pellegrino2} considering a viscous system in the presence of a magnetic field, which provides additional possibilities to study magnetohydrodynamics.

Experiments on $PdCoO_{2}$ \cite{moll}, $WP_{2}$ \cite{gooth},
and GaAs \cite{molenkamp, gusev, gusev2, levin} have many features demonstrating the viscous flow of electrons.
Moreover, the previous study of the giant negative magnetoresistance in high mobility GaAs structure \cite{haug, hatke, mani, haug2}
could be interpreted as a manifestation of the viscosity effects, or interplay
between ballistic and hydrodynamic effects \cite{alekseev2}.

\begin{figure}[ht]
\includegraphics[width=\linewidth]{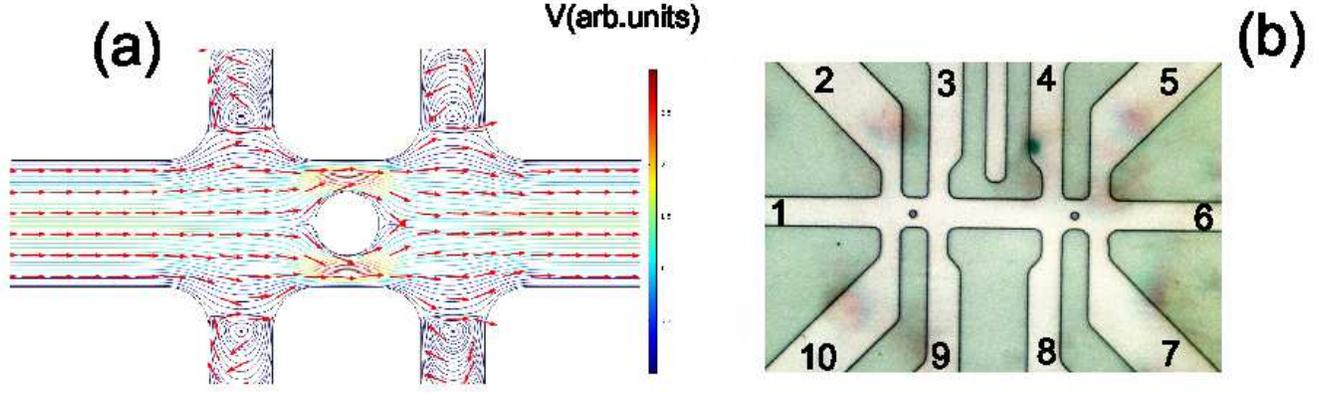}
\caption{(Color online) (a) Sketch of the velocity flow profile in the presence of a circular obstacle. (b) Image of the Hall bar device with
antidot (micro-hole) in the center of the Hall bridge between probes 2-3 (10-9) and 4-5 (8-7).}\label{fig:1}
\end{figure}

The diffusive scattering condition is the relevant one for most liquid-solid
interfaces. The absence of Poiseuille flow and the Gurzhi effect in
graphene has been taken as evidence for a specular limit with
a very large slip length \cite{bandurin}.

If the slip length is larger than sample size, viscous shear forces can arise, if the fluid flows around an obstacle.
 Flow around a circular disc was considered
by Stokes a long time ago \cite{stokes, landau}. In classical two-dimensional
fluid mechanics, this may lead to a phenomenon referred as the "Stokes paradox":  no solution of the Stokes equations can be found for which fluid velocity
satisfies both the boundary conditions on the body and at infinity \cite{lamb}.

Recently an electronic analog of the Stokes paradox has been proposed
for two-dimensional Fermi liquids \cite{hruska, guo, lucas1}. Schematically this proposal is illustrated in Figure \ref{fig:1}a: the resistance of the sample with length $L\sim W$
is studied, when a circle obstacle of radius $a_{0} << L$ is located in the middle of the sample \cite{lucas2, kisilev}. In an electronic liquid, the Stokes paradox
has been resolved within the framework of the semiclassical description of quasiparticle dynamics, and a linear response has been obtained due to the momentum relaxation process \cite{hruska, guo, lucas2}.
Indeed Ohmic theory predicts that the obstacle will enhance total resistance \cite{lucas2} :
\begin{equation}
R_{total}=R_{0}+R_{obst},
\end{equation}
where $R_{0}$ is obstacle free resistance, and
$R_{obst}=cR_{0}\frac{a_{0}^{2}}{L^{2}}$, c is a geometric factor.
It is interesting that the Stokes flow around a disc leads to a dramatic consequence beyond Ohmic behaviour: the effective radius of the obstacle $a_{eff}$
is always larger than the geometric radius $a_{eff} >> a_{0}$ \cite{lucas2}. More importantly the obstacle resistance decreases with temperature, suggesting that the viscous liquid is essentially always in the regime of specular scattering boundary
conditions.

In the present work, we have experimentally examined the transport properties of a mesoscopic 2D electron system with a circular obstacle (antidot or micro-hole).
As a reference we also studied a device without an antidot in order to extract the obstacle resistance and determine all relevant viscous parameters,
which provides the comparative analysis between theory and experiment. By tuning the temperature in a wide interval $ 1.5 < T < 70 K$, we show that
obstacle resistance $R_{obst}$ exhibits a drop as temperature increases (even as $dR_{0}/dT > O$), in consistence with predictions for the ballistic and hydrodynamic regimes.

\section{Methods}

The samples were grown by molecular
beam epitaxy method. Our samples are high-quality, GaAs quantum wells with a width of 14~nm
with electron density $n_{s}=6\times10^{11} cm ^{-2}$ and a mobility of $\mu=2.5\times10^{6} cm^{2}/Vs$ at T=1.4K.
Other parameters, such as fermi velocity, mean free path and others are given in Table \ref{tab1}.
We present experimental results on Hall-bar devices.
They consist of three, $6 \mu m$ wide segments of
different length ($6, 20 , 6 \mu m$), and 10 contacts. Figure \ref{fig:1}b shows the image of a typical multiprobe Hall device I.
The antidots are located in the middle of the right side and left side segment of the Hall bar by chemical wet etching through the quantum well.
The measurements were carried out in a
VTI cryostat, using a
lock-in technique to measure the longitudinal $\rho_{xx}$ resistivity with an
ac current of $0.1 - 1 \mu A$ through the sample.
3 Hall bars from the same wafers were studied and showed consistent behaviour. As reference we also measured a Hall bar
without an antidot. Additionally we also studied macroscopic samples, where, it is expected, that the viscous effects are small. These samples have Hall-bar geometry
(length $l\times$ width $W = 500 \mu m \times 200 \mu m$) with  six  contacts.

\begin{table}[ht]
\centering
\begin{tabular}{|l|l|l|l||l||l||l|}
\hline
&W&$n_{s}$ & $v_{F}$ & $l$  & $l_{2}$ & $\eta$\\
&($\mu m$) & $(10^{11} cm^{2}$) & $(10^{7} cm/s)$ &  $(\mu m$) & $(\mu m$) & $(m^{2}/s)$\\
\hline
&$6$& $6.0$  & $3.3$ &  $35$ & $3$ & $0.25$\\
\hline
\end{tabular}
\caption{\label{tab1} Parameters of the electron system  at $T=1.4 K$. Parameters $l$, $l_{2}$ and $\eta$ are determined in the text.}
\end{table}

\section{Experiment in reference device and discussion}

\begin{figure}[ht]
\includegraphics[width=\linewidth]{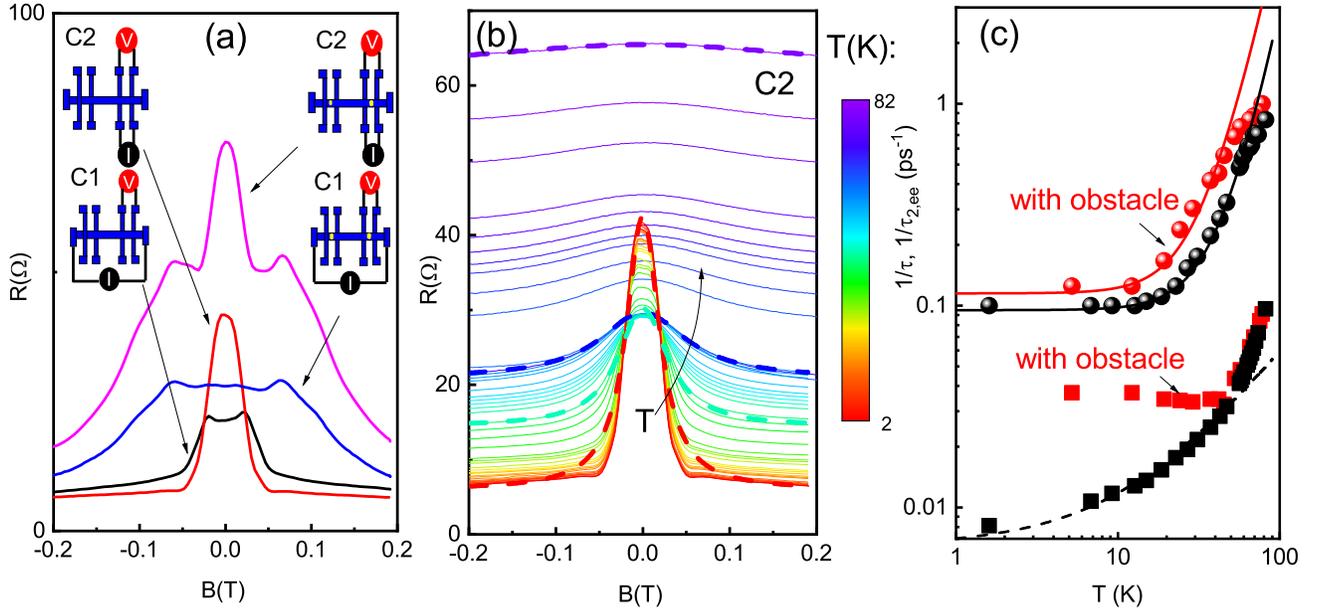}
\caption{(Color online)
(a) The magnetoresistance of a GaAs quantum well in a Hall bar sample with obstacle and in a reference sample for different configurations, T=1.4K. The schematics
show how the current source and the voltmeter are connected for the measurements:
configuration with antidot is shown on the right side, configuration for reference without antidot is shown on the left side.
(b) Temperature dependent magnetoresistance of a reference Hall bar sample. Dashes are examples illustrating
magnetoresistance calculated from Eq. 2 for different temperatures: 1.5 K (blue), 27.7 K(light blue), 44 K (green).
(c) Relaxation rates $1/\tau$ (squares) and  $1/\tau_{2,ee}$ (circles) as a function of the temperature obtained by fitting
the theory with experimental results for the reference sample (black scatters) and a sample with an obstacle (red scatters).
Thick black and red lines is Equation 3, dashes is Equation 4.
}\label{fig:2}
\end{figure}

The electronic analog of the hydrodynamic regime in the pipe is a electric current in a narrow channel of width $W\sim 1-10\mu m$.
Figure \ref{fig:1}b shows the image of the Hall bar device with a micro-hole in the center of the Hall bridge. The resistance between different probes
has been measured.
  Figure \ref{fig:2}a shows the longitudinal magnetoresistance for a sample with an antidot and a reference sample without an antidot.
Longitudinal magnetoresistance of a viscous 2D high mobility system in GaAs has been studied in previous research for different configurations of current and voltage probes \cite{gusev, gusev2, levin}.
Remarkably, we find that probe configuration and sample geometry strongly affect the temperature evolution of
local resistance and its value at zero magnetic field. For example, when the current is applied between probes 1 and 6, and voltage is measured between probes 4 and 5 (referred further as C1 configuration),
the corresponding resistance $R_{I=1-6;V=4-5}$ increases with temperature T, while the resistance $R_{I=8-7;V=4-5}$, when the current is applied between probes 8 and 7 and voltage is measured between probes 4 and 5 (referred further as C2 configuration), decreases with T and always appears bigger
than $R_{I=1-6;V=4-5}$. We attribute such behaviour to enhanced viscosity due to diffusive scattering on the rough edge
and inhomogeneity of the velocity field, predicted in paper \cite{alekseev}. Indeed we reproduced these results in the samples studied in this work, and Figure \ref{fig:2}a  shows that
the resistance at B=0 in configuration C2 is bigger than the resistance in configuration C1. Moreover, the resistance with an antidot is enhanced
in comparison with the reference sample in both configurations. One more striking feature is the anomalously large negative magnetoresistance, which is strongly enhanced for configuration C2.
Satellite peaks are clearly observed in samples with antidots resulting in additional broadening of the total magnetoresistance.
Therefore, we may conclude here that the effect of the obstacle is adding a series resistor, as has been predicted in paper \cite{lucas2}.
Before analyzing the obstacle effect, and in order to make this analysis more complete, we present the results of measurements of longitudinal magnetoresistivity $\rho_{xx}(B)$ in samples without a micro-hole.
In order to increase the viscosity effect, we study resistance in C2 configuration.
Figure \ref{fig:2}b shows $\rho_{xx}(B)$ as a function of magnetic field and temperature.

In the hydrodynamic approach, the semiclassical treatment of the transport describes the motion of carriers when the higher order moments of the distribution function are taken into account.
The momentum relaxation rate $1/\tau$ is determined by electron interaction with phonons and static defects (boundary).
The second moment relaxation rate $1/\tau_{2,ee}$ leads to the viscosity and contains the contribution
from the electron-electron scattering and temperature independent scattering by disorder \cite{alekseev, scaffidi}.
It has been shown that conductivity is determined by two independent $\textit{parallel}$ channels of electron momentum relaxation:
the first is due to momentum relaxation time and the second due to viscosity \cite{alekseev, scaffidi}.
This approach allows the introduction of the magnetic field dependent viscosity tensor and the derivation of the magnetoresisivity tensor \cite{alekseev, scaffidi, luca}:
\begin{equation}
\rho_{xx}= \rho_{0}^{bulk}\left(1+\frac{\tau}{\tau^{*}}\frac{1}{1+(2\omega_{c}\tau_{2,ee})^{2}}\right),\,\,\,
\end{equation}

where $\rho_{0}^{bulk}=m/ne^2\tau$, $\tau^{*}=\frac{W(W+6l_{s})}{12\eta}$, viscosity $\eta=\frac{1}{4}v_{F}^{2}\tau_{2,ee}$.

All relaxation rates are given by:
\begin{equation}
\frac{1}{\tau_{2,ee}(T)}=A_{ee}^{FL}\frac{T^{2}}{[ln(E_{F}/T)]^{2}}+\frac{1}{\tau_{2,0}},
\end{equation}

where $E_{F}$ is the Fermi energy, and the coefficient $A_{ee}^{FL}$  be can expressed via the Landau interaction parameter.
The relaxation rate $\frac{1}{\tau_{2,0}}$ is not related to the electron-electron collisions, since any process responsible for relaxation of the second moment
of the distribution function, even scattering by static defect, gives rise to viscosity \cite{alekseev}.
The momentum relaxation rate is expressed as:
\begin{equation}
\frac{1}{\tau}=A_{ph}T+\frac{1}{\tau_{0}},
\end{equation}
where $A_{ph}$ is the term responsible for the phonon scattering, and $\frac{1}{\tau_{0}}$ is the scattering rate due to static disorder
(not related to the second moment relaxation rate $\frac{1}{\tau_{2,0}}$). It is worth noting that above 40 K the scattering from polar LO
phonons becomes important and the scattering time deviates from simple linear dependence on temperature \cite{haris, kawamura}).

We fit the magnetoresistance curves in Figure \ref{fig:2}b  and the resistance in zero magnetic field with the 3 fitting parameters :
$\tau(T)$, $\tau^{*}(T)$ and $\tau_{2,ee}(T)$.
We compare the temperature dependence of  $\frac{1}{\tau_{2,ee}(T)}$ and $\frac{1}{\tau(T)}$ with theoretical predictions given by Equations 3 and 4,
which is shown in Figure \ref{fig:2}c.
The following parameters
are extracted:
$1/\tau_{2,0}=0.95\times10^{11}$ s, $A_{ee}^{FL}=0.35\times10^{9} s^{-1}K^{-2}$,
$A_{ph}=0.5\times10^{9}sK^{-1}$ and $1/\tau_{0}=0.65\times10^{10} s$,
which are correlated with previous studies \cite{gusev,levin}. Note, however, that a discrepancy with Equations 3 and 4 is found at high temperatures,
which we attributed to the inelastic process due to scattering by LO phonons .
Relaxation time  $\tau^{*}(T)$ depends on  $\tau_{2,ee}(T)$ and the boundary slip length $l_{s}$.
Comparing these values, we find that $l_{s}=3.2 \mu m < L$. Our data are in good agreement with the theoretical prediction for the case
when the slip length is temperature independent.
Table 1 shows the mean free paths : $l=v_{F}\tau$, $l_{2}=v_{F}\tau_{2,ee}$  and viscosity, calculated with parameters extracted from the fit of experimental data.

In the last part of this section, we wish to discuss the influence of the ballistic effect on negative magnetoresistance in our reference samples.
As we already mentioned in the introduction, a previous study of the magnetoresistance in high mobility two dimensional GaAs system
demonstrated giant two-scale negative magnetoresistance consisting of a narrow temperature independent peak near zero magnetic field and
 shoulder-like magnetoresistance, which strongly depends on the temperature \cite{haug2}. The model \cite{alekseev2} proposes, that the temperature independent
  peak is attributed to the ballistic effects, while shoulder is attributed to the hydrodynamic effects due to flowing between randomly located
  macroscopic “oval” defects. It is worth noting that, because we observe small size peaks in magnetoresistance in C1 configuration (Figure \ref{fig:2}a),
  ballistic contribution, predicted in the model \cite{alekseev2} can have a non-negligible effect at least at low temperature. We present two  arguments justifying, that ballistic effect
  is smaller than hydrodynamic contribution. First, we have demonstrated that magnetoresistance and $R(T)$ strongly depend on the configuration ( C1 or C2),
  which is unlikely to be attributed to the ballistic effect \cite{gusev, gusev2, levin}. For example, ballistic contribution can not describe
  the resistance drop with temperature ( Gurzhi effect), observed in our samples \cite{gusev}. Second,
  our giant negative magnetoresitance strongly depends on temperature and can be successfully described  within a hydrodynamic framework \cite{alekseev}
  in wide temperature range, in contrast to the T-independent peak observed in paper \cite{haug2}. However, even though both ballistic and hydrodynamic
  contribution are equally important at low temperature, at high temperature, the viscosity effect becomes dominant, and all our conclusion can be applied
  equally well to the samples with and without obstacle.

\section{Experiment: obstacle resistance}

\begin{figure}[ht]
\includegraphics[width=\linewidth]{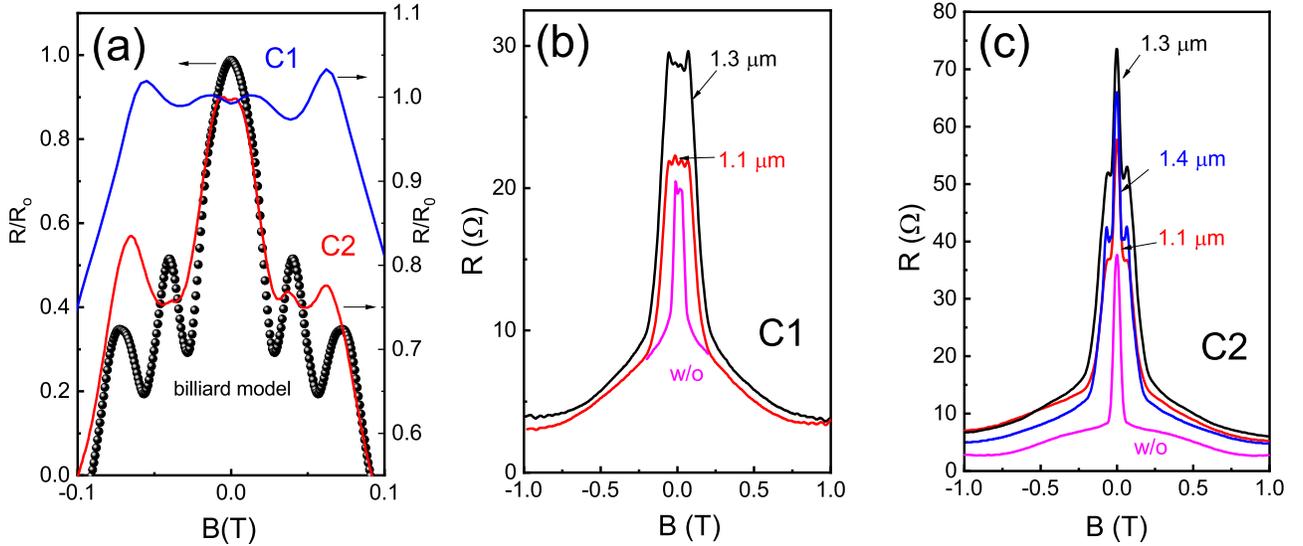}
\caption{(Color online)
(a) Magnetoresistance for a sample with an obstacle in C1 and C2 configurations, T=4.2K. The dots represent results for the billiard model.
The magnetoresistance of samples with different obstacle radii and in the reference sample (without obstacle) for configurations C1 (b) and C2 (c), T=4.2K.}\label{fig:3}
\end{figure}

In this section, we focus on the study of resistance in samples with an obstacle. Figure \ref{fig:3}a shows the magnetoresistance for samples with an
obstacle for both configurations C1 and C2. One can see small satellite peaks making the central peak wider in comparison with the reference sample.
We attribute these oscillations to geometrical resonance effects, which are pronounced in 2D charged liquids \cite{beenakker, roukes}.
We perform numerical simulations of the electron trajectories in ballistic structures for different obstacle sizes (for details see Supplementary material).
The results of theses simulations (dots) for $a_{0}=1 \mu m$ are compared to the experimental data. One can see, that the width of the magnetoresistance
curve roughly corresponds to the experimental data, while the position of the peak is slightly shifted to a higher
magnetic field in comparison with the experiment.

Magnetoresistance as a function of the magnetic field for different radii $a_{0}$ is shown in Figures \ref{fig:3}b,c for two configurations C1 and C2.
The diameter of the antidot has been measured directly from an optical microscope image (Figure \ref{fig:1}b) with precision $0.1\mu m$. The effective
antidot diameter is larger than the lithographic one due to the depletion region, which, however, in our high density sample does not exceed $0.05\mu m$.
We estimate this value from the assumption that the width of the region where the potential increases from the bottom to the Fermi energy is of the same order as the Fermi
wavelength for typical electron concentrations \cite{ando}.
Traces for the reference sample without an obstacle are shown for comparison. One can see that the resistance
with an obstacle is always larger than the reference resistance. Resistance of a sample with an antidot radius of $a_{0}=1.3 \mu m$ is higher than the resistance with $a_{0}=1.4 \mu m$, probably due to radius uncertainty ($\pm 0.05 \mu m$). Viscosity effects are enhanced in C2 configurations and below
we focus on the results obtained from this probe configuration.

\begin{figure}[ht]
\includegraphics[width=\linewidth]{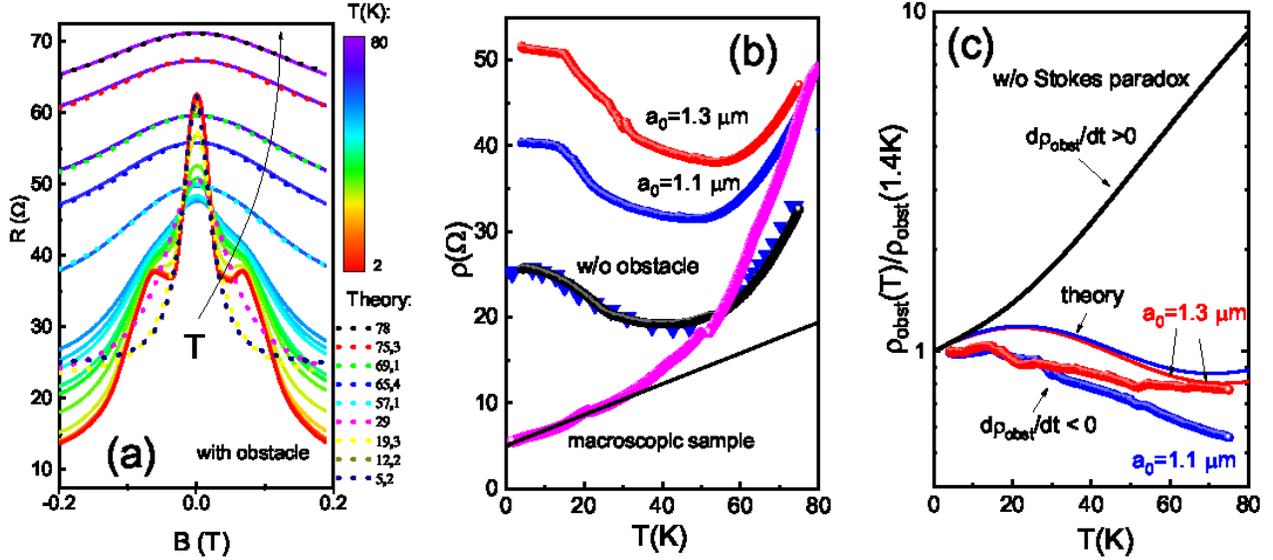}
\caption{(Color online)
(a) Temperature dependent magnetoresistance of a sample with obstacle ($a_{0}=1.3 \mu m$). Dashes is
magnetoresistance calculated from Eq. 1 for  4.2 K with parameters taken from fit with the reference sample's magnetoresistance.
(b) Temperature dependent resistivity of a sample with an obstacle, reference sample and macroscopic sample. Triangulares are
resistivity calculated from Eq. 2. Solid line represents resistivity due to acoustic phonon scattering. (c) Relative obstacle resitivities
for samples with different obstacle radii. Colors solid line represents calculations from Equation 8 with numerical parameters taken from magnetoresistance
measurements in the reference sample. Black solid line represents calculations without the Stokes paradox effect. Obstacle resistance exhibits
a drop with decreasing temperature ($d\rho_{obst}/dT < 0$).} \label{fig:4}
\end{figure}

Figure \ref{fig:4}a shows the evolution of magnetoresistance with temperature for samples with an obstacle in C2 configuration.
We fit a central peak with the Lorentzian curve (Eq.2). Note that this peak is absent in magnetoresistance for C1 configuration (Figure \ref{fig:2}a and Figure \ref{fig:3}a)
because it is overlapped by satellite peaks. As for the reference sample, we used the 3 fitting parameters :
$\tau(T)$, $\tau^{*}(T)$ and $\tau_{2,ee}(T)$.
Figure \ref{fig:2}c shows the relaxation rates $1/\tau(T)$, and $1/\tau_{2,ee}(T)$ for an obstacle sample in comparison with the reference sample as a function of temperature.
One can see that the rate $1/\tau_{2,ee}(T)$ is following the dependencies of Eqs. 3-4 with parameters
$1/\tau_{2,0}=1.15\times10^{11}$ s, $A_{ee}^{FL}=0.9\times10^{9} s^{-1}K^{-2}$, while the rate $1/\tau(T)$ is
saturated at low temperatures, and it is unlikely that it can be described by the acoustic phonon scattering mechanism. The difference between
rates $1/\tau_{2,ee}(T)$ for obstacle and reference samples can be attributed to uncertainty in the determination of the Lorentz curve
width due to the satellite ballistic peak. The momentum relaxation rate is extracted from resistivity at zero magnetic field, which is enhanced
in the obstacle samples.

The temperature dependence of resistivity at zero magnetic field for different obstacle radii and the reference sample in configuration C2 is shown in Figure \ref{fig:4}b.
Note, that for our approximately square-shaped devices (Figure \ref{fig:1}b), 2D resistivity practically equals the resistance: $R=1.6\rho$, and below we discuss the resistivity behaviour.
One can see that resistance (resistivity) decreases in the temperature interval $1.5 K < T< 45 K $ and increases at higher temperatures.
We argue here that the ballistic (quasiballistic) contribution is described by the first term Equation 2, and
comparison with theory proves that it is much smaller than the viscosity contribution described by the second term.  Below we repeat several keyword arguments which
justify this conclusion and which have been discussed in previous publications \cite{gusev, levin}. First, the resistivity for C2 configuration decreases with temperature and follows the Gurzhi law $\rho\sim T^{-2}$
at least at low T (see Figure \ref{fig:2}c) \cite{gusev}. In contrast, resistivity in macroscopic samples increases with T and follow the linear law $\rho\sim T$ (below 40 K), due to acoustic phonon scattering (see
Figure \ref{fig:4}b) \cite{haris, kawamura}.
Therefore, we would expect that resistivity due to moment relaxation is temperature independent (scattering with static defects or boundary) or increases with T (due to the phonon scattering mechanism).
Second, the electron-electron scattering obeys the power law $\frac{1}{\tau_{2,ee}(T)}\sim T^{2}$ (the logarithmic term is weakly T-dependent) \cite{gusev, levin}, instead of the linear T law
expected for phonon scattering. We compared the experimental dependence of $\rho(T)$  in zero magnetic field with theoretical models and
obtained a good agreement (see Figure \ref{fig:4}b -triangulares). Finally, resistivity strongly depends on the probe configuration (Figure \ref{fig:2}a), which is unlikely to be attributable to the ballistic effect.
Indeed,  we calculated the ballistic contribution
in our sample geometry and found only weak dependence on the configuration, which disagrees with our observations.

In the Figure \ref{fig:4}b, we can see that resistivity of the samples with obstacles is always larger than the resistivity of the reference sample
within the investigated temperature range. The enhanced obstacle resistivity $\rho_{obst}(T)=\rho_{total}(T)-\rho_{0}(T)$ as a function of temperature
is shown in Fig. \ref{fig:4}c for two obstacle radii. For comparison we demonstrate the resistivity measured in a macroscopic sample $\rho_{macr}$. Conventional Ohmic behaviour
is expected in this device:
below 40 K,
macroscopic resistivity displays simple linear temperature dependence due to acoustic phonon scattering (shown by solid line), while at higher temperatures
scattering from polar LO phonons starts to become important. Indeed $d\rho_{macr}/dT >0$ in the entire interval of temperatures. In contrast
obstacle resistance shows $d\rho_{obst}/dT < 0$ in the same temperature region.

\section{Theory and discussion}

Simplified Ohmic theory predicts that obstacle resistivity should be proportional to obstacle free resistance and the square of the
obstacle radius \cite{lucas} $\rho_{obst}(T) \sim \rho_{0}(T)(\frac{a_{0}}{L})^{2}$. Therefore, one might expect that obstacle resistivity just
reproduces the temperature dependence of the Ohmic resistivity. The solid line in Figure \ref{fig:4}c represents the resulting
obstacle resistivity without viscosity effects, when only phonon scattering (acoustic and LO phonons) is
taken into account. It predicts a very strong ($\sim 10$ times) increase of  $R_{obst}(T)$ with temperature, which indeed
disagrees with our experiments. One may conclude here that the T- coefficient of obstacle resistance is attributed to the combination
of two effects: viscous flow of electrons in a narrow sample and the hydrodynamic flow around the obstacle.

As we already mentioned in the introduction, a lot of theoretical effort has gone into the resolution of the Stokes paradox
in two-dimensional charged liquids. The main result is that the effective radius of the obstacle is larger
than the geometric radius $a_{0}$ and depends on temperature. The inverse scattering length drastically affects electron flow
behaviour in the presence of an obstacle: $\frac{1}{l_{eff}}=\frac{1}{l}+\frac{1}{l_{2}}$

Three different regimes of the transport have been considered \cite{lucas2}:

(i) Diffusive: in this limit $a_{0} >> \sqrt{l_{eff}l_{2}}$, and effective radius is give by
\begin{equation}
a_{eff}=a_{0}.
\end{equation}

(ii) Ballistic: in this limit $l_{eff} >> a_{0}$, and effective radius is give by
\begin{equation}
 a_{eff}^{2}=\frac{a_{0}l_{2}}{2}.
\end{equation}

(iii) Hydrodynamic : in this limit

$l_{eff} << a_{0}<< \sqrt{l_{eff}l_{2}}$, and effective radius is give by
\begin{equation}
 a_{eff}^{2}=\frac{l_{eff}l_{2}}{ln(\frac{l_{eff}l_{2}}{a_{0}^{2}})}.
\end{equation}

\begin{figure}[ht]
\includegraphics[width=\linewidth]{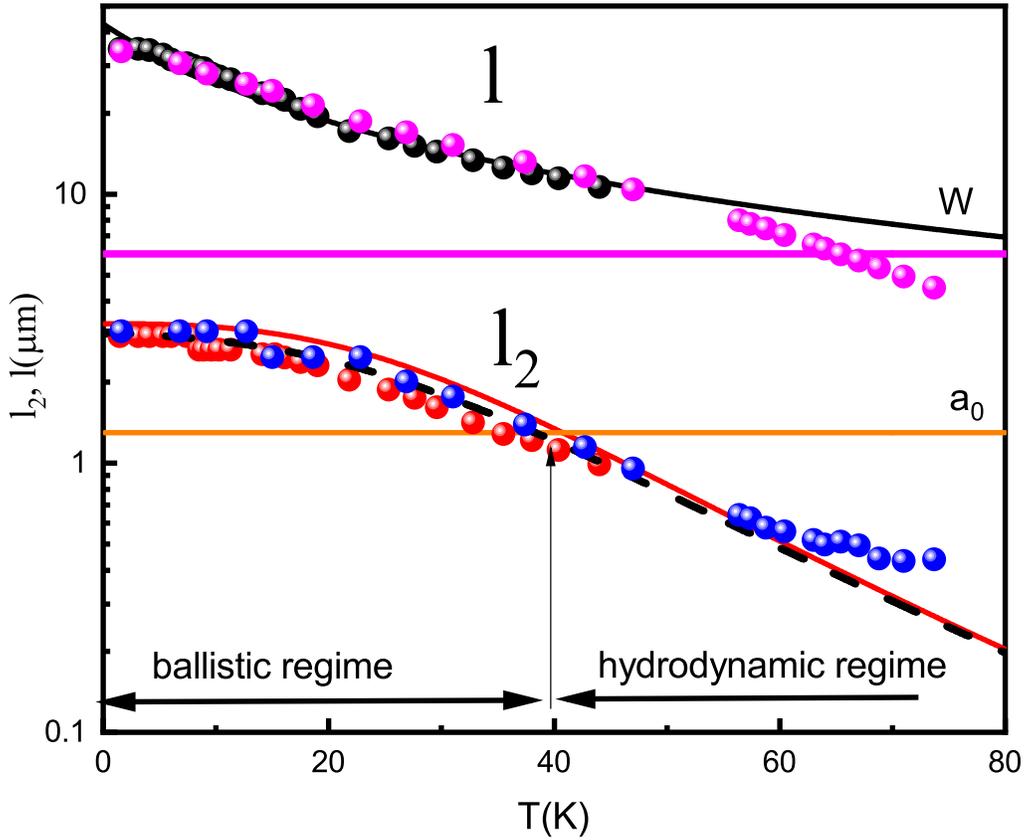}
\caption{(Color online)
The characteristic lengths $l$, $l_{2}$ and  $l_{eff}$ (dashes) as a function of temperature.
Dots-parameters obtained from magnetoresistance measurements in the two reference samples. Fit of characteristic length
with parameters indicated in the main text. Horizontal lines- the width of the sample $W$ and
diameter of the obstacle $a_{0}$. Ballistic and hydrodynamic regimes on the length scale of the disk are shown.}\label{fig:5}
\end{figure}

This difference in the parameter regimes leads to markedly different physical behavior in the transport.
It is remarkable that, in the hydrodynamic regime, the effective radius only weakly depends on the actual radius $a_{0}$.
In order to compare our results with theoretical predictions for corresponding transport limits, we
calculate relevant electron parameters as a function of temperature. Figure \ref{fig:5} represents temperature
dependence of the characteristic lengths $l$, $l_{2,ee}$ and $l_{eff}$ extracted from experiments on the two reference samples.
One can see that the viscous regime conditions $l_{2,ee} < W < l$ are satisfied in all temperature intervals, which is justified
by observation of the Gurzhi effect below $T < 40 K$. Since obstacle radius is much smaller than the width of the sample,
the hydrodynamic limit for the Stokes effect requires higher temperatures $T > 40 K$. Model \cite{lucas2} predicts
a general behavior for the effective obstacle radius, which covers all transport regimes:

\begin{eqnarray}
a_{eff}^{2}\approx l_{eff} l_{2}\Bigg\{\left(1-\frac{2l_{eff}}{l_{2}}\right)\times \nonumber \\
 \times\log \left[\frac{l_{2}}{l_{eff}}\left(\sqrt{1+\left(\frac{2l_{eff}}{a_{0}}\right)^{2}}-1\right)+1\right] \nonumber \\
 + \sqrt{1+\left(\frac{2l_{eff}}{a_{0}}\right)^{2}}-1 \Bigg\}^{-1}.
\end{eqnarray}

We compared the prediction of this model with our results, which are shown in Figure \ref{fig:4}c. The theory predicts
slightly nonmonotonic behaviour of $\rho_{obs}(T)$ due to the interplay between $\rho_{0}(T)$ and $a_{eff}(T)$ dependencies:
at low temperatures, contribution from obstacle free resistivity is dominant, while at higher temperatures, the effective radius exhibits
a sharp drop due to viscosity. We can see that the predicted results roughly agree with experimental observations due to the approximate character of the
analytical calculations. It is because the theory \cite{lucas2} does not consider collisions with the sample boundary, which lead to a quadratic velocity
profile in the sample and a viscous character of the flow even without an obstacle.

It is important to note that $d\rho_{obst}/dT < 0$ in the whole temperature interval, which disagrees
with macroscopic resistivity behavior ($d\rho_{macr}/dT >0$) and mesoscopic total resistivity behaviour (with and without antidots), displaying nonmonotonic behaviour :
$d\rho_{total}/dT <0$ for $1.4 K < T < 40 K $ and  $d\rho_{total}/dT > 0$ for $ 40 K < T < 80 K$ .

\section{Summary and conclusion}

We have studied an electronic analog of the Stokes flow around the obstacle in a two-dimensional system in high quality GaAs quantum wells.
The resistance of 2D electrons with a micro-hole fabricated in the center of the sample is always enhanced in comparison
with obstacle-free devices. Obstacle resistance decreases with temperature even as $d\rho_{0}/dT >0$. Experimental results confirm the theoretically predicted significance
of momentum relaxation in the ballistic and hydrodynamic regimes, which is significantly distinct from conventional Ohmic behaviour.

\section{Acknowledgment}
We thank J.P.Peña for her help in sample fabrication. The financial support of this work by FAPESP (Brazil),
CNPq (Brazil) and  the Russian Science
Foundation (Grant No.16-12-10041) is acknowledged.

\section*{Author contributions statement}

G. M. G., A. D. L., Z. D. K and A. S. J. performed the experiment,
 A. K. B. synthesized the crystals, G. M. G.,
and A. D. L. provided the theoretical framework, G. M. G. wrote
the manuscript with inputs from all authors. G. M. G  and Z. D. K.
supervised the work.  All authors reviewed the manuscript.

\section*{Additional information}

Supplementary information accompanies this paper at https://.
To include, in this order: \textbf{Accession codes} (where applicable); \textbf{Competing financial interests} (The authors declare that they have no competing interests).

The corresponding author is responsible for submitting a \href{http://www.nature.com/srep/policies/index.html#competing}{competing financial interests statement} on behalf of all authors of the paper.

\end{document}